\documentclass[
  journal=pasa,
  manuscript=research-article,
  year=2026,
  volume=XX
]{cup-journal}
\usepackage[nopatch]{microtype}
\usepackage{booktabs}
\usepackage{graphicx}%
\usepackage{multirow}%
\usepackage{amsmath,amssymb,amsfonts}%
\usepackage{amsthm}%
\usepackage{mathrsfs}%
\usepackage[title]{appendix}%
\usepackage{xcolor}%
\usepackage{textcomp}%
\usepackage{manyfoot}%
\usepackage{booktabs}%
\usepackage{breakcites} 
\usepackage{listings}%
\usepackage[nolist]{acronym}

\usepackage[colorlinks=true,linkcolor=blue]{hyperref}
\hypersetup{citecolor=blue, urlcolor=blue}

\begin{acronym}[AWGN]
\acro{ASKAP}{Australian Square Kilometre Array Pathfinder}
\acro{ARC}{Australian Research Council}
\acro{ATNF}{Australia Telescope National Facility}
\acro{CASDA}{CSIRO ASKAP Science Data Archive}
\acro{CSIRO}{Australian Commonwealth Scientific and Industrial Research Organisation}
\acro{EMU}{Evolutionary Map of the Universe}
\acro{IR}{infrared}
\acro{IRSA}{Infrared Science Archive}
\acro{NIR}{Near infrared}
\acro{BPT}{Baldwin, Phillips \& Terlevich}
\acro{RMS}[rms]{root mean squared}
\acro{SN}{supernova}
\acro{SNR}{supernova remnant}
\acro{AGN}{active galactic nucleus}
\acro{SMBH}{Super Massive Black Hole}
\acro{RACS}{Rapid ASKAP Continuum Survey}
\acro{HIPASS}{H{\sc i} Parkes All-Sky Survey}
\acro{WISE}{Wide-field Infrared Survey Explorer}
\acro{eROSITA}{extended Roentgen Survey with an Imaging Telescope Array}
\acro{eRASS}{eROSITA All-Sky Survey}
\acro{LMC}{Large Magellanic Cloud}
\acro{NRAO}{National Radio Astronomy Observatory}
\acro{ATCA}{Australia Telescope Compact Array}
\acro{WR}{Wolf-Rayet}
\acro{CABB}{Compact Array Broadband Backend}
\acro{DECaPS2}{The Dark Energy Camera Plane Survey 2}
\acro{CABB}{Compact Array Broadband Backend}
\acro{ULX}{ultraluminous X-ray source}
\acro{DECam}{The Dark Energy Camera}
\acro{MC}{Magellanic Cloud}
\acro{SMBH}{supermassive black hole}
\acro{SN}{supernova}
\acro{QSO}{Quasi Stellar Objects}
\acro{SFMS}{Star Formation Main Sequence}
\acro{ISM}{interstellar medium}
\acro{CARTA}{Cube Analysis and Rendering Tool for
Astronomy}
\acro{FWHM}{Full Width at Half Maximum}
\acro{DRAGN}{Double Radio-source Associated with Galactic Nucleus}
\end{acronym}

\newcommand{\ngc}{NGC\,5938}

\def\arcsec{\hbox{$^{\prime\prime}$}}

\title{EMU Radio Observations of Barred Spiral Galaxy NGC\,5938 (Araish)}

\author{H. Zakir}

\affiliation{School of Science, Western Sydney University, Locked Bag 1797, Kingswood, 2751, NSW, Australia}
\email[Hina Zakir]{22110114@student.westernsydney.edu.au}

\author{M.D. Filipovi\'c}
\affiliation{School of Science, Western Sydney University, Locked Bag 1797, Kingswood, 2751, NSW, Australia}

\author{L. Barnes}
\affiliation{School of Science, Western Sydney University, Locked Bag 1797, Kingswood, 2751, NSW, Australia}

\author{R. Z. E. Alsaberi}
\affiliation{Faculty of Engineering, Gifu University, 1-1 Yanagido, Gifu 501-1193, Japan}
\alsoaffiliation{School of Science, Western Sydney University, Locked Bag 1797, Kingswood, 2751, NSW, Australia}

\author{T. An}
\affiliation{Shanghai Astronomical Observatory, CAS, 80 Nandan Road, Shanghai 200030, P.R. China} 
\alsoaffiliation{State Key Laboratory of Radio Astronomy and Technology, A20 Datun Road, Chaoyang District, Beijing, P. R. China} 

\author{K. Dage}
\affiliation{Curtin Institute of Radio Astronomy, Curtin University: Perth, Western Australia, AU} 

\author{S. W. Duchesne}
\affiliation{Australia Telescope National Facility, CSIRO, Space and Astronomy, PO Box 1130, Bentley, WA 6151, Australia}  

\author{A.~M. Hopkins}
\affiliation{School of Mathematical and Physical Sciences, Macquarie University, 12 Wally’s Walk, Macquarie Park, 2109, NSW, Australia}

\author{A. Kapinska}
\affiliation{National Radio Astronomy Observatory (NRAO), 1011 Lopezville Rd, Socorro NM 87801, USA}

\author{B.~Koribalski}
\affiliation{Australia Telescope National Facility, CSIRO, Space and Astronomy, PO Box 76, Epping, NSW 1710, Australia}
\alsoaffiliation{School of Science, Western Sydney University, Locked Bag 1797, Kingswood, 2751, NSW, Australia}

\author{S. Lazarevi\'c}
\affiliation{School of Science, Western Sydney University, Locked Bag 1797, Kingswood, 2751, NSW, Australia}
\alsoaffiliation{Australia Telescope National Facility, CSIRO, Space and Astronomy, PO Box 76, Epping, NSW 1710, Australia}
\alsoaffiliation{Astronomical Observatory, Volgina 7, 11060 Belgrade, Serbia}

\author{D. Leahy}
\affiliation{Department of Physics and Astronomy, University of Calgary, Calgary, Alberta, T2N 1N4, Canada}

\author{Z. Liu}
\affiliation{Max Planck Institute for extraterrestrial Physics, Giessenbachstr. 1, 85748 Garching, Germany}

\author{R.~P.~Norris}
\affiliation{Australia Telescope National Facility, CSIRO, Space and Astronomy, PO Box 76, Epping, NSW 1710, Australia}
\alsoaffiliation{School of Science, Western Sydney University, Locked Bag 1797, Kingswood, 2751, NSW, Australia}

\author{A. Rau}
\affiliation{Max Planck Institute for extraterrestrial Physics, Giessenbachstr. 1, 85748 Garching, Germany}

\author{Z.J. Smeaton}
\affiliation{School of Science, Western Sydney University, Locked Bag 1797, Kingswood, 2751, NSW, Australia}

\author{T. Jarrett$^*$}
\affiliation{Department of Astronomy, University of Cape Town, Private Bag X3, Rondebosch 7701, South Africa}
\alsoaffiliation{School of Science, Western Sydney University, Locked Bag 1797, Kingswood, 2751, NSW, Australia}
\alsoaffiliation{$^*$Deceased on 3~July~2024}


\begin{document}

\begin{abstract}

We present multi-wavelength observations of the nearby spiral galaxy \ngc\ (Araish) to investigate the origin of its radio emission, specifically the contribution from \ac{AGN} activity and star formation. Using \ac{EMU} data, we detect extended radio emission extending outwards to the galactic axis, with a steep non-thermal spectral index ($\alpha = -1.2 \pm 0.2$) indicative of synchrotron radiation from an \ac{AGN} jet. The jet has a physical extent of $\approx 8.2$\,kpc (angular length of 64\arcsec). Multi-wavelength data from \ac{DECaPS2}, \ac{WISE}, and \ac{eROSITA} provide further support for this interpretation. The colour-colour diagram presenting \ac{WISE} infrared observations suggests the presence of dust and young stars that trace the galaxy's disk structure. Our analysis reveals a radio jet, alongside star formation traced by infrared emission, demonstrating the complex interplay of \ac{AGN} activity and star formation in this well-resolved galaxy. 
Intriguingly, the spatial relationship reveals the brighter X-ray emission to be largely adjacent to and enveloping the extended radio emission. This suggests that the radio jet, while extending at a significant angle to the galactic disk, is confined by the larger X-ray gas/halo, similar to other systems (i.e., ESO~295--IG022, Centaurus~A) and may indicate jet collimation and channelling effects.\footnote{\thanks{Accepted for publication in PASA.}}

\end{abstract}

\acresetall

\begin{figure*}[!ht]
\centering
\includegraphics[width=1.0\linewidth]{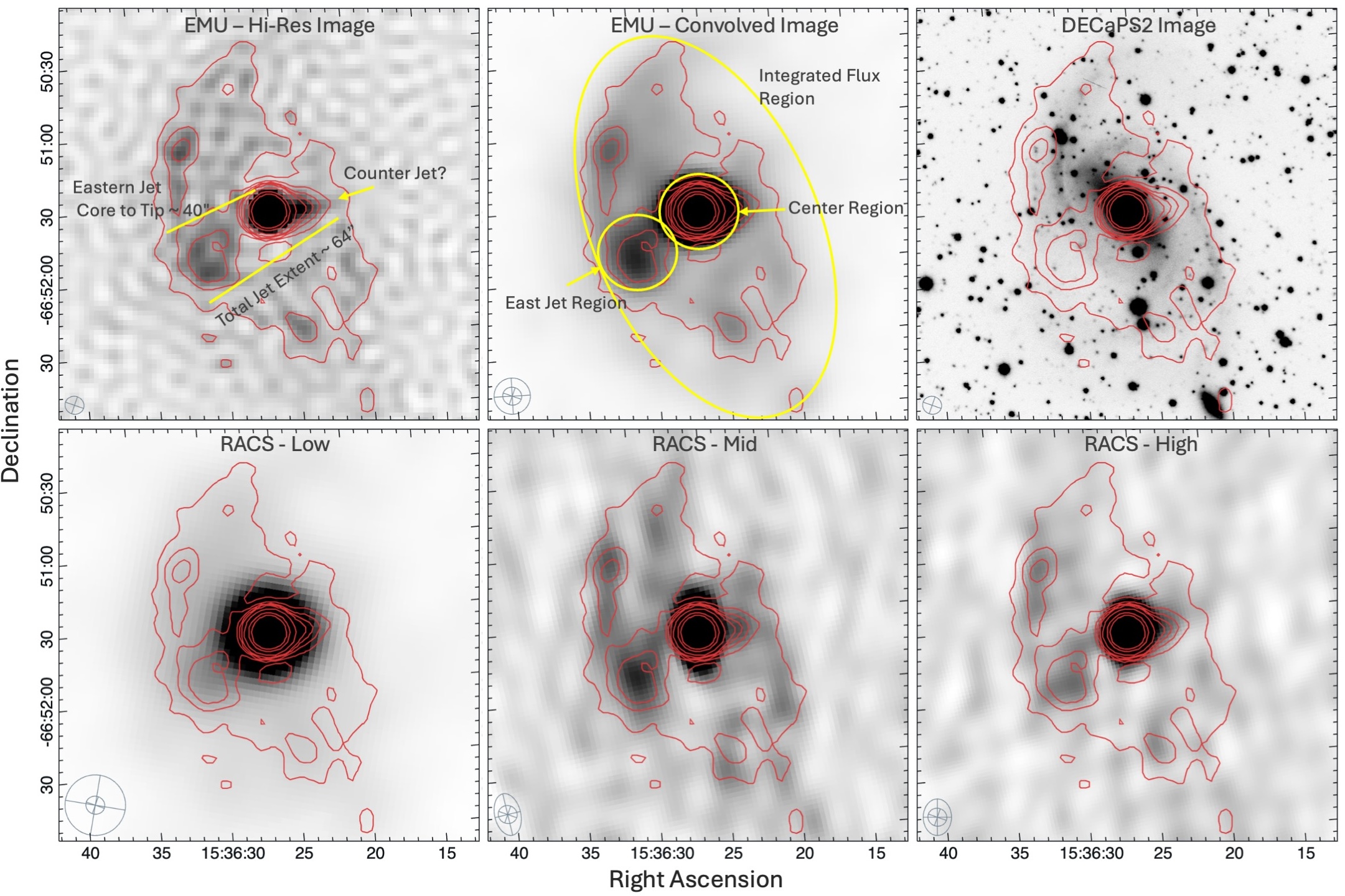}
\caption{Radio and NIR observations of \ngc\ (Araish). The top row (from left to right) shows the EMU Hi-Res image, the EMU Convolved Image, and the DECaPS2 optical z-band image. The bottom row shows the RACS Low, RACS Mid, and RACS High band images with the EMU Hi-Res contours overlaid at 3, 6, 9, 12, 15, 25, 50, and 75\,$\sigma$  on all panels, where $\sigma$ = $65\mu$Jy\,beam$^{-1}$.}
\label{fig_0}
\end{figure*}

\section{Introduction}

Radio emission in galaxies can originate from multiple astrophysical processes, primarily driven by compact objects such as \acp{SNR}, massive star-forming regions, or central \acp{SMBH} producing \ac{AGN}. In particular, \ac{AGN}-driven jets emit strongly at radio wavelengths via synchrotron radiation and can significantly influence galaxy evolution by regulating star formation through feedback processes \citep{2012MNRAS.421.1569B, 2012ARA&A..50..455F}. However, disentangling \ac{AGN} activity from star formation in galaxies, especially at average resolutions, remains challenging, as both can contribute substantially to the total radio output \citep{2007MNRAS.375..931M}.

While powerful radio jets are commonly observed in elliptical galaxies or massive quasars, their presence in spiral galaxies is rare. These systems, often referred to as spiral \aclu{DRAGN} \citep[\ac{DRAGN};][]{10.1093/mnrasl/sly081, 10.1093/mnras/stv2071, Norris2025}, represent a unique class where classical disk morphology coexists with large-scale radio jets \citep{Gao_2023}. Spiral \ac{DRAGN}s challenge traditional paradigms of \ac{AGN}-hosting environments and raise important questions about jet triggering mechanisms, gas accretion histories, and the role of galaxy mergers in otherwise secularly evolving systems.

Notable examples include 0313-192 \citep{Ledlow_2001}, a massive edge-on spiral galaxy with a radio jet of $\approx 300$ kpc, and NGC 3079, a barred spiral hosting a compact radio outflow and kpc-scale X-ray bubble \citep{Cecil_2001}. The Circinus Galaxy \citep{1998MNRAS.297.1202E}, although not formally classified as a \ac{DRAGN}, hosts Seyfert-like nuclear activity, a starburst ring, and collimated radio structures. These systems serve as critical laboratories for probing the interplay between star formation, \ac{AGN} feedback, and \ac{ISM} dynamics in non-elliptical hosts.

In this paper, we present new multi-wavelength observations of \ngc, an edge-on spiral galaxy exhibiting prominent radio emission suggestive of \ac{AGN}-driven activity. Observations were obtained as part of the \aclu{EMU} \citep[\ac{EMU}; ][]{2021_Norris, hopkins2025evolutionarymapuniversenew} survey conducted with the \aclu{ASKAP} \citep[\ac{ASKAP}; ][]{2021_Hotan}. Despite its regular spiral appearance, it shows extended radio emission over a projected extent of $\sim33\arcsec$ ($\approx 4.2$ kpc), consistent with structures commonly associated with \ac{AGN} activity \citep{10.1093/mnras/stx403}. Among the possible mechanisms responsible for this emission are an extended radio jet or a compact radio lobe, similar to those observed in nearby systems such as NGC 3079 \citep{Cecil_2001}, where a kpc-scale radio outflow emerges from the disk.

\begin{sloppypar}
We also use complementary observations from the \aclu{WISE} \citep[\ac{WISE}; ][]{ 2010AJ....140.1868W}, \aclu{DECaPS2} \citep[\ac{DECaPS2}; ][]{Saydjari_2023}, and the \aclu{eROSITA} \citep[\ac{eROSITA}; ][]{Predehl_2021} to examine the interplay between the radio jet and star formation, as traced by infrared and X-ray emission. The spatial morphology and multi-wavelength signatures of \ngc\ place it among the rare class of spiral galaxies with \ac{AGN}-driven outflows, as visible in NGC 1068 \citep{Young_2001, Cecil2002}.
\end{sloppypar}

\ngc\ is nicknamed \textit{Araish}, a word meaning ``adornment'' in Urdu, reflecting the galaxy’s striking spiral morphology and active star-forming disk. This visual impression, particularly the representation of its orderly disk and the emerging radio jet, motivated the informal use of this name in the manuscript.

A detailed characterisation of the target galaxy is provided in Section~\ref{S:target}. The observational data are described in Section~\ref{S:DataObs}, followed by an analysis of the jet and host morphology in Section~\ref{RadioDiscussion}. Our findings are described in the context of similar systems in Section~\ref{S:Results} and we conclude with a summary of the implications for spiral galaxies exhibiting both \ac{AGN} and star-formation signatures.

\section{Target Description: NGC 5938 (Araish)}
\label{S:target}

\ngc\ is a nearby barred spiral galaxy located at RA = 15$^\mathrm{h}$36$^\mathrm{m}$18$^\mathrm{s}$, Dec = $-$66$^\circ$13$'$47$''$ (J2000), and is catalogued in multiple surveys, including the \ac{HIPASS}, as HIPASS~J1536$-$66 \citep{2004AJ....128...16K}. It exhibits a well-defined disk morphology with B, R, and I band magnitudes of 11.9, 12.0, and 11.5, respectively \citep{2005MNRAS.361...34D}. Despite its regular spiral appearance, it shows extended radio emission consistent with a radio lobe or jet. Such features are rare among spiral galaxies and are commonly classified as spiral \ac{DRAGN}s, a class that challenges the traditional view that large-scale radio jets occur only in elliptical hosts \citep{Gao_2023}. These galaxies serve as valuable laboratories for studying the interplay of \ac{AGN} feedback and disk galaxy evolution.

\ngc\ has a redshift of $z = 0.012$, corresponding to a systemic H\textsc{i} velocity of $v_\mathrm{sys} = 3512 \pm 9$\,km\,s$^{-1}$ \citep{2004AJ....128...16K}. In the Local Group frame, the velocity becomes $v_\mathrm{LG}=3314$~km\,s$^{-1}$. A simple application of the Hubble–Le maître law with $H_0 = 75$\,km\,s$^{-1}$\,Mpc$^{-1}$ yields a Hubble-flow distance of $\sim\,$44.2\,Mpc. However, redshift-based distances are uncertain at such low redshift values due to large peculiar motions \citep{2002A&A...393...57T}.

To derive a more accurate distance, we apply redshift-independent methods. The \textit{sosie} galaxy technique \citep{2007A&A...465...71T} and the Tully–Fisher relation \citep{2020AJ....160..199S} both yield consistent values of $\sim\,$26.6\,$\pm$\,0.4\,Mpc and $\sim\,$26\,Mpc, respectively. We adopt the \textit{sosie} galaxy distance of 26.6\,Mpc for all physical calculations in this work.

\ngc\ contains a substantial neutral hydrogen reservoir with H\textsc{i} flux of $F_\mathrm{HI} = 26.4 \pm 3.7$\,Jy\,km\,s$^{-1}$ and a derived H\textsc{i} mass of $\log(M_\mathrm{HI}/M_\odot) \approx 10.27$ \citep{2004MNRAS.350.1195M}. The H\textsc{i}-WISE survey \citep{Parkash_2018} further estimates a stellar mass of $\log(M_\ast/M_\odot) \approx 11.2$ and star formation rate of $\log(\mathrm{SFR}/M_\odot\,\mathrm{yr}^{-1}) \approx 0.58$, placing it near the star-forming main sequence \citep{Whitaker2012}.

\begin{table*}[hbt!]
    \caption{EMU and RACS radio survey flux density and spectral index measurements of regions within NGC 5938.}
    \label{tab_Rflux}
    \begin{tabular}{cccc}
    \toprule
     Frequency$^{a}$  &   East Jet ($33\arcsec\approx4.2$\,kpc) & Centre ($32\arcsec\approx4.1$\,kpc)  & Galaxy-Integrated ($176\arcsec\approx22.7$\,kpc)  \\
      (MHz)          &    (mJy)   & (mJy) & (mJy)  \\
    \midrule
    888$^{\rm L}$  & $4.3 \pm 0.46$  & $24.6 \pm 0.44$ &  \textcolor{gray!70}{$81.9 \pm 2.44$} \\
    944$^{\rm E}$  & $4.4 \pm 0.09$ & $34.1 \pm 0.09$ & $90.6 \pm 0.49$ \\
    1368$^{\rm M}$ & $3.0 \pm 0.60$  & $21.0 \pm 0.59$ &  \textcolor{gray!70}{$49.2 \pm 3.85$} \\
    1655$^{\rm H}$ & $2.0 \pm 0.69$  & $18.9 \pm 0.67$ &  \textcolor{gray!70}{$31.8 \pm 3.66$} \\
    \hline
    Spectral Index $\alpha$ & $-1.2 \pm 0.2$ & $-0.7 \pm 0.4$ &  \textcolor{gray!70}{$-1.5 \pm 0.2$} \\
    \hline
    888 MHz - Luminosity(W/Hz) & $3.7\times10^{20}$ & $2.9\times10^{21}$ &  \textcolor{gray!70}{$7.0\times10^{21}$ }\\
    1000 MHz - Luminosity(W/Hz)  & $3.4\times10^{20}$ & $2.2\times10^{21}$ &  \textcolor{gray!70}{$6.2\times10^{21}$ }\\
    \midrule
    \bottomrule
    \end{tabular}
    \begin{tablenotes}
    \scriptsize
    \item $^{a}$ RMS and beam values for each survey: $^{\rm L}$ RACS Low (888 MHz) rms = 0.25 mJy/beam, 15$^{\prime\prime}$ beam; $^{\rm E}$ EMU (944 MHz) rms = 0.05 mJy/beam, 15$^{\prime\prime}$ beam; $^{\rm M}$ RACS Mid (1368 MHz) rms = 0.25 mJy/beam, 12$^{\prime\prime}$ beam; $^{\rm H}$ RACS High (1655 MHz) rms = 0.25 mJy/beam, 10$^{\prime\prime}$ beam. Greyed-out values mark RACS flux densities that may be underestimated due to the higher rms noise and lower angular resolution of the RACS data. Flux uncertainties are calculated as $\delta S = \mathrm{rms} \sqrt{N_{\rm beam}}$.
    \end{tablenotes}

\end{table*}


\begin{figure*}
\centering
\includegraphics[width=\linewidth]{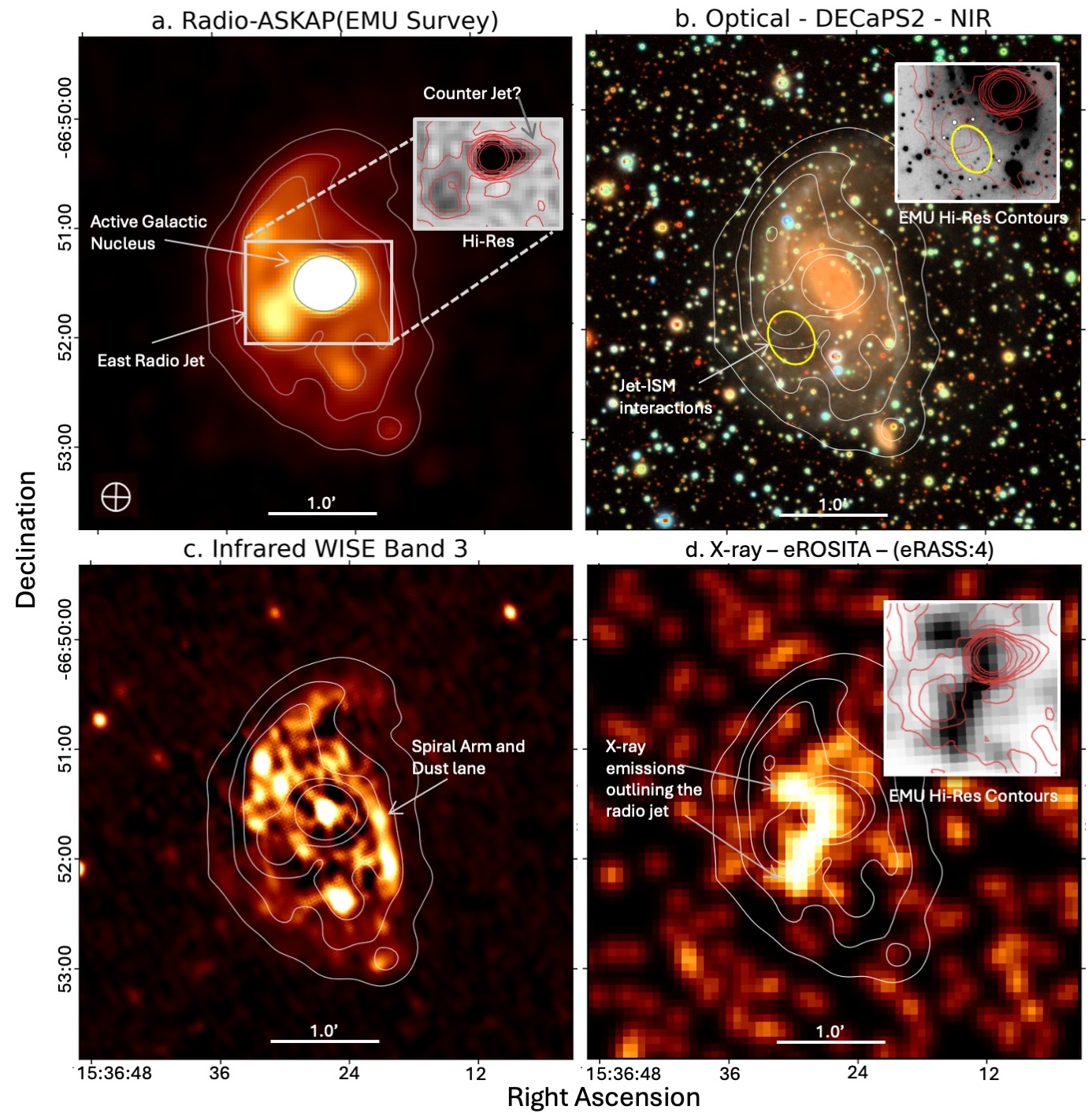}
\caption{Multi-wavelength observations of the spiral galaxy \ngc\ (Araish) a: Radio continuum image from \ac{EMU} 944\,MHz tile SB62461, Convolved Image has the beam size of $15\arcsec$, at PA\,=\,0$^{\circ}$, with an EMU Hi-Res inset image showing finer structures of the nucleus.
b: DECaPS2 optical/NIR GRZ (475\,nm, 644\,nm, 926\,nm) bands composite image represented by Blue, Green and Red colours respectively, with inset showing Hi-Res radio contours overlaid on the DECaPS2 z-band image, highlighting jet–ISM cavity overlap.
c: \acs{WISE} Band-3. 
d: eROSITA X-ray Observation: 0.2--5.0\,keV, Gaussian smoothing at \ac{FWHM}=$3\arcsec$, with inset showing Hi-Res radio contours overlaid on X-ray image, highlighting radio jet and X-ray adjoining.
The EMU convolved radio contours are applied on all base images at 3, 9, 15, 25, and 50\,$\sigma$, where $\sigma$ = $75\mu$Jy\,beam$^{-1}$. The radio emission from the jet region is encompassed by the $25\sigma$ radio contour.}
\label{fig_mul}
\end{figure*}

\section{Data and Observations} 
 \label{S:DataObs}

\subsection{Radio: EMU}

The radio continuum morphology of \ngc\ as observed with the \ac{EMU} survey \citep{hopkins2025evolutionarymapuniversenew, 2021_Norris, 2021_Hotan} is presented in Figure~\ref{fig_0}. The observation was conducted on 21$^{\rm st}$ May~2024 using the full complement of 36 \ac{ASKAP} antennas, at the central frequency of 943\,MHz with a 288\,MHz bandwidth (scheduling blocks SB62461). The data were processed using the ASKAPsoft pipelines, including multi-frequency synthesis imaging and multi-scale cleaning. For a complementary view and quantitative analysis of the overall emission, the data were also convolved to a common beam size of 15\arcsec$\times$15\arcsec\ with a position angle of PA = $0$ degrees, from which flux densities  were calculated. The higher resolution \ac{EMU}-HiRES image, as shown in Figure~\ref{fig_0}, was utilised to effectively illustrate the detailed radio structure of \ngc\ and to study the centre of the galaxy for the existence of a potential counter jet. Hi-Res processing enhances the visibility of fine details within galaxies, \ac{AGN}, and other radio sources, as demonstrated by \citet{Guzman_2019}.

Additionally, we detected \ngc\ in the \aclu{RACS} \citep[\ac{RACS};][]{2020PASA...37...48M, Hale_2021, 2023PASA...40...34D, 2025PASA...42...38D} across its Low ($\nu \approx 888~\mathrm{MHz}$), Mid ($\nu \approx 1367~\mathrm{MHz}$), and High ($\nu \approx 1655~\mathrm{MHz}$) frequency bands. RACS provides wide-field continuum imaging with angular resolutions of approximately $15\arcsec$ (Low), $12\arcsec$ (Mid), and $10\arcsec$ (High), and typical sensitivities of $\sim250$--$300~\mu\mathrm{Jy~beam^{-1}}$. Measured integrated flux densities for each band are listed in Table~\ref{tab_Rflux}. In all three \ac{RACS} bands, \ngc\ is detected as a bright, compact radio core coincident with the optical nucleus, accompanied by extended emission associated with the eastern jet region (spans $\sim33\arcsec$). This extended component has integrated flux densities of $4.3\pm0.46$~mJy (888~MHz / Low band), $3.0\pm0.60$~mJy (1368~MHz / Mid band), and $2.0\pm0.69$~mJy (1655~MHz / High band). While the Low band exhibits the highest integrated flux density, consistent with a non-thermal spectrum, the emission morphology is most clearly resolved and detected in the RACS-Mid band (Figure~\ref{fig_0}). This difference in visual clarity is attributed to the varying angular resolution across the bands. The Low band has the largest beam, which spreads the emission, resulting in lower peak surface brightness and a less distinct morphology, despite having the highest total flux.
The detected surface-brightness levels are low, corresponding to only $1.2 - 1.6$ times the \ac{RACS} rms noise ($\sim0.25$~mJy~beam$^{-1}$). It is important to note that the \ac{RACS} sensitivity of $\sim 0.25~\mathrm{mJy~beam^{-1}}$ (rms) is roughly five times lower than the \ac{EMU} survey sensitivity of $\sim 0.05\,\text{mJy beam}^{-1}$. This difference may lead to an underestimation of flux density in \ac{RACS}, particularly in regions of low surface brightness where faint emission approaches the detection threshold.

Flux densities for the \ac{EMU} convolved image (SB62461) and the three \ac{RACS} bands were measured for the jet region, the galaxy centre, and the integrated flux region using \aclu{CARTA} \citep[\ac{CARTA}; ][]{2021ascl.soft03031C}. We used \ac{CARTA}’s statistical tools to measure the integrated flux density within the elliptical apertures shown in the EMU Convolved Image Figure~\ref{fig_0}. These apertures were manually defined to enclose the contiguous emission above the local $6\sigma$ level for the east jet and the central region, while the galaxy-integrated region encompasses all emission above $3\sigma$. The semi-major axes of these apertures correspond to the characteristic angular extents listed in Table \ref{tab_Rflux}.
\begin{itemize}
 \item Central region: an ellipse with semi-major axis $\sim32\arcsec$ centred on the optical nucleus
 \item East Jet region: an ellipse with semi-major axis $\sim 33\arcsec$, centred on the peak of the eastern extension, encloses the jet-like emission 
 \item Galaxy Integrated region: a larger ellipse with semi-major axis $\sim 176\arcsec$ encloses the full galaxy-integrated emission above $3\sigma$ level.

\end{itemize}

Flux density values were extracted using the image statistics panel, and background noise was estimated locally to subtract any baseline offset. No primary beam correction was applied. Flux uncertainties were estimated using the rms noise expressed in mJy\,beam$^{-1}$, following the standard relation for uncorrelated noise per synthesised beam:
\[
\delta S = \mathrm{rms}\times\sqrt{N_{\mathrm{beam}}}
\]

where $N_{beam}$ is the number of synthesised beams within the flux–extraction region,
\[
N_{\mathrm{beam}} = A_{\mathrm{flux}} / A_{\mathrm{beam}}
\]
and $A_{beam}$ is the solid angle area of the synthesized beam.
\[ A_{\mathrm{beam}}=\pi\theta_{\mathrm{maj}}\theta_{\mathrm{min}}/(4\ln 2)\]

This provides the correct noise propagation when integrating flux across extended emission. The measured flux densities and corresponding uncertainties are reported in Table \ref{tab_Rflux}, where greyed-out values represent RACS measurements for which the flux density may be underestimated due to the lower angular resolution and higher rms noise of the RACS data, particularly in regions of low surface brightness where emission approaches the detection threshold. We used these flux density measurements to calculate the spectral index ($\alpha$)\footnote{The spectral index ($\alpha$) is the logarithmic slope of flux density vs. frequency ($S_\nu \propto \nu^{\alpha}$).} and the radio luminosity given in (Table~\ref{tab_Rflux}).

The EMU in-band spectral index of the galaxy's nucleus were also measured using multi-frequency synthesis imaging with Taylor terms $T_0$ and $T_1$. The EMU Taylor images are outputs of the multi-frequency synthesis process, representing the sky brightness as a Taylor series expansion as a function of frequency. The $T_0$ term corresponds to the intensity image at the reference frequency, while $T_1$ represents the first-order spectral slope component. The spectral index $\alpha$ for the centre region was derived from the ratio 

\[
\alpha = \frac{T_1}{T_0},
\]
following the approach outlined by \citet{Rau2011}. The flux density  of the centre region $T_0$ was measured as $3.72 \times 10^{-2}$\,Jy with a standard deviation of $7.71 \times 10^{-3}$\,Jy beam$^{-1}$, while $T_1$ was $-2.61 \times 10^{-2}$\,Jy with a standard deviation of $5.64 \times 10^{-3}$\,Jy beam$^{-1}$. This results in a spectral index $\alpha = -0.70 \pm 0.21$, indicating a steep spectral slope characteristic of non-thermal emission processes. This result is consistent with the value derived from EMU and RACS flux densities (Table~\ref{tab_Rflux}). The spectral index is further discussed in Section \ref{RadioDiscussion}. All data are available through the \ac{ASKAP} Science Data Archive (CASDA\footnote{\url{https://research.csiro.au/casda}}).

 \begin{figure*}[hbt!]
     \centering
     \includegraphics[width=0.8\linewidth]{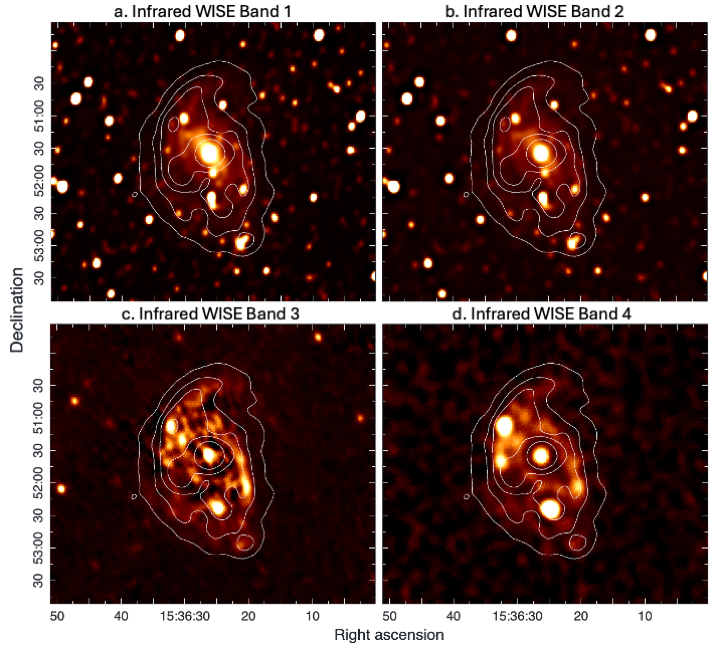}
     \caption{Infrared observations of \ngc\, a: \ac{WISE} Band W1, b: Band W2, c: Band W3, d: Band W4 images. 3, 9, 15, 25, and 50\,$\sigma$ contours applied from \ac{EMU} convolved radio observations, where $\sigma$ = $75\mu$Jy\,beam$^{-1}$. }
     \label{fig_wiseall}
 \end{figure*}

\subsection{Optical/NIR: DECaPS2}
Figure~\ref{fig_mul}b shows \ngc\ in optical and near-IR images from \ac{DECaPS2} \citep{Saydjari_2023}\footnote{We use the Legacy Survey Sky browser \url{https://decaps.legacysurvey.org/}}. The single-exposure limiting magnitudes reach $(23.5, 22.6, 21.6)$ mag in the GRZ (475\,nm, 644\,nm, 926\,nm) bands.

\subsection{IR: WISE}

 Infrared observations (Figures~\ref{fig_mul}c and \ref{fig_wiseall}) were obtained from \ac{WISE} \citep{2022RNAAS...6..188M} where band W1, W2, W3, and W4 correspond to 3.4\,$\mu$m, 4.6\,$\mu$m, 12\,$\mu$m, and 22\,$\mu$m respectively. Image cutouts were retrieved from Caltech datasets available through the \ac{IRSA} at NASA/IPAC\footnote{\url{https://irsa.ipac.caltech.edu/frontpage/}}. The \ac{WISE} AllWISE release provides typical $5\sigma$ point-source sensitivities of 16.5, 15.5, 11.2 and 7.9 mag (Vega) in W1-W4, which correspond to approximate flux density limits of 0.08, 0.11, 1.0 and 6.0 mJy. These sensitivity levels are sufficient to detect faint stellar populations and dust emission in nearby galaxies, although sensitivity degrades in regions of high background noise such as the Galactic plane \citep{2010AJ....140.1868W}. We also calculated the infrared Vega magnitudes across W1-W4 images as listed in Table \ref{tab_wise}.

\subsection{X-Ray: eROSITA}
Figure~\ref{fig_mul}d shows the X-ray image constructed from the combined data from the first four All-Sky Surveys eRASS \citep{Merloni_2024}, eRASS1, eRASS2, eRASS3, and eRASS4, obtained with the \ac{eROSITA} \citep{Predehl_2021} onboard the Russian-German SRG mission \citep{Sunyaev_2021}. The eROSITA data were calibrated and cleaned using the pipeline version 020 of the eROSITA Science Analysis Software (eSASS, \citealt{brunner_etal2022}, version \texttt{20211004}). We merged the photon events from all seven telescopes into one event list file for each eRASS. To increase the signal-to-noise ratio, we also generated an event list combining all the data from the four eRASS epochs (hereafter eRASS:4). 
We applied eRASS:4 (0.2--5.0\,keV) image smoothing using a Gaussian kernel with a \ac{FWHM} of 3\arcsec\ using the Python Astropy 2D convolution kernel functions. This smoothing scale optimally reveals the features of interest, specifically the X-ray cocoon surrounding the radio emission.

\section{Results and Discussion} 
 \label{S:Results}
    
\subsection{Radio Observations}\label{RadioDiscussion}

As shown in Figure~\ref{fig_0}, \ac{EMU} Hi-Res and Convolved observations, there is an extended region of radio emission perpendicular to the galaxy's major axis, resembling possible \ac{AGN} jets originating from the central \ac{SMBH}. Taking the redshift independent distance of $\approx 26.6$\,Mpc to the galaxy, angular measurements indicate the core to tip length of eastern jet $\approx 40\arcsec$ and the total jet extent of the \ac{AGN} $ \approx 64\arcsec$, which converts into physical sizes $ \approx 5.15$\,kpc and $ \approx 8.2$\,kpc, respectively. Notably, Figure~\ref{fig_mul}a (Hi-Res Inset image) tentatively labels a feature to the north-west of the nucleus as a ``counter jet''. This feature appears to be approximately aligned with the East jet axis but displaced from a direct line of sight, potentially indicating that it is angled out of the plane. 
Interpretation of this feature include a counter (west) jet, or possibly backflow.

Similar to the well-studied and more energetic galaxy NGC\,1068 \citep{Young_2001}, which hosts prominent \ac{AGN} jets, our observations indicate that the jet is oriented roughly perpendicular to the galactic disk and its major axis. However, this does not definitively establish the jet’s 3D orientation relative to the galaxy’s actual disk. The jet could be truly perpendicular to the disk, extending directly out of the plane, or it could be inclined at some angle, like in NGC\,1068. It is worth noting that in NGC\,1068, despite the active \ac{AGN}, the star formation contributes significantly to the radio emission and the \ac{AGN} itself accounts for only about half the total radio luminosity \citep{Cecil2002}. If the jet in \ngc\ is similarly inclined, it may still be propagating through the outer regions of the \ac{ISM}, rather than through the main disk itself.

Evidence for the jet-ISM interaction, which could help distinguish between these scenarios \citep{10.3389/fspas.2017.00042}, is limited in our current data. The optical image (Figure~\ref{fig_mul}b) shows morphological features that are consistent with jet-ISM interaction, though the infrared image (Figure~\ref{fig_mul}c) show no sign of obscuration of the dust lanes, nor any obvious heating of gas or dust along the jet's path. These observations suggest that the jet is likely propagating through relatively low-density ISM, producing subtle structural effects (e.g. a shallow cavity) but without strong radiative or thermal signatures typical of jet impact on dense molecular gas.

The central \ac{AGN} region has a spectral index of $\alpha =  - 0.7\pm0.4$, while the eastern jet region shows a steeper index of $\alpha = -1.2\pm0.2$. This steeper index suggests a non-thermal emission mechanism, consistent with synchrotron radiation \citep{book2}. The spectral index of \ngc's centre was also calculated using the ratio of \ac{EMU} Taylor T0 and T1 images; This method gives us the spectral index $\alpha = -0.7\pm0.2$, which is consistent with the spectral index calculated using the flux densities from \ac{EMU} and RACS surveys (Table~\ref{tab_Rflux}). The observation of radio emission in \ngc\ extending perpendicular to the galaxy's major axis (Figure~\ref{fig_mul}a), combined with the non-thermal spectral index, strongly supports the interpretation of an \ac{AGN} jet.

\subsection{Optical Observations}
 \label{subsec2}

The optical observations from \ac{DECaPS2} (Figure~\ref{fig_mul}b) show the galaxy in visible light, highlighting the luminous nucleus, distribution of stars and the overall morphology. \ngc\ is classified as a barred spiral, with a smooth disk and a less prominent central bulge compared to most spiral galaxies, following \cite{Moorthy_2006}. The dust lanes are apparent in the optical, partially obscuring the galaxy's optical emissions from the nucleus. The \ac{DECaPS2} z-band image (Figure~\ref{fig_mul}b) reveals a diffuse stellar and ISM distribution, within which a pronounced under-luminous cavity (highlighted by the yellow ellipse) is identifiable. This cavity appears to be caused by a low density of stars or dust absorption in the region. Although the origin of this cavity is uncertain, it is spatially aligned with the radio-jet axis ($f_{\mathrm{overlap}} \allowbreak \approx 50\%$, white contours). This geometric association suggests a possible imprint of the jet on the surrounding ISM \citep{10.3389/fspas.2017.00042}. Such a marginal structural depression would be consistent with a jet propagating through comparatively low-density medium, producing subtle clearing without the strong heating or dust-lane disruption typically associated with jet–ISM interaction in denser regions of galactic disks.

\begin{figure}[ht!]
  \centering
   \includegraphics[width=\columnwidth]{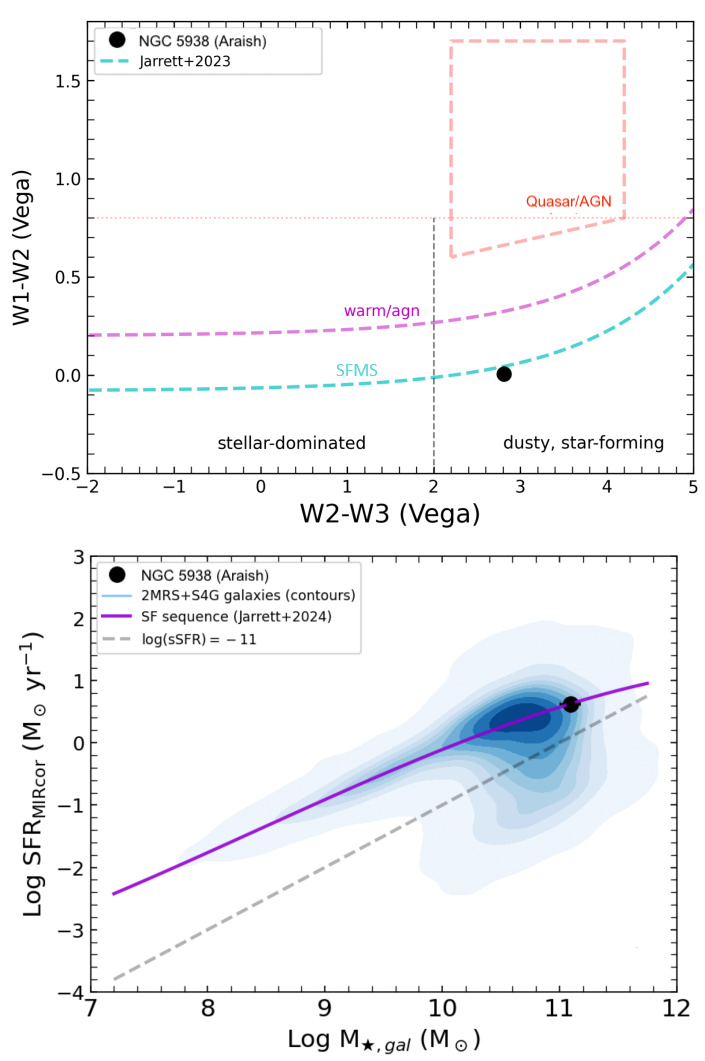}
    \caption{Top: \ac{WISE} colour-colour diagram showing \ngc's position relative to stellar-dominated, dusty star-forming, SFMS, warm/AGN, and Quasar/AGN regions. The dashed curves mark template tracks from \cite{Jarrett_2023}. The location of \ngc\ (black dot) near the dusty star-forming / warm–AGN boundary indicates the presence of significant dust. Bottom: Star Formation Main Sequence diagram plotting mid-infrared–corrected star formation rate (SFR$_{\mathrm{MIRcor}}$) derived from WISE 12 $\mu$m luminosity following \cite{2024MNRAS.529.3210B} versus stellar mass. The contours represent the distribution of nearby galaxies from the combined 2MRS+S4G sample from \cite{Jarrett_2023}, while the solid magenta line shows the SFMS fit from \cite{2024MNRAS.529.3210B}. The dashed grey line corresponds to a specific star formation rate of log(sSFR) = –11 yr$^{-1}$. \ngc\ located on the SFMS, consistent with active star formation.}
   \label{fig_SF}
\end{figure} 
  
\subsection{Infrared Observations}
 \label{subsec3}

Unlike visible light, infrared emission can pass through dust, making it invaluable for studying obscured star formation and \ac{AGN}. The longer wavelengths of infrared radiation enable us to detect star formation and \ac{AGN} that are invisible due to dust blockage in the optical spectrum. It also traces the distribution and properties of the dust itself, a key component of the \ac{ISM} \citep{Saydjari_2023}. \ngc\ observations from \ac{WISE} \citep{2010AJ....140.1868W} W1-W4 bands (Figure~\ref{fig_wiseall}, Table~\ref{tab_wise}) show prominent infrared emission . 


In the W1 band, the galaxy exhibits an apparent magnitude of \( 8.10 \pm 0.02 \). For context, \citet{10.1093/mnras/stu1087} shows that \ac{WISE} sources with \( \mathrm{W1} < 11 \) are typically dominated by stars, and extragalactic objects become more prevalent at fainter magnitudes, i.e., \( \mathrm{W1} > 11 \)\footnote{\url{https://faculty.washington.edu/ivezic/Publications/WISE1.pdf}}. Therefore, our target’s W1 magnitude of \( 8.10 \) places it firmly among the brightest extragalactic objects detected in W1, suggesting a substantial 
near-infrared luminosity.

Moreover, mid-infrared studies of \ac{AGN} populations, such as those utilising \ac{WISE} often focus on sources with W1 magnitudes in the range $10\text{--}15\,\mathrm{mag}$ i.e.\,\ac{AGN} selection criteria by \citet{Stren2012} and \citet{OConnor2016}. 
Within these samples, an object with \( \mathrm{W1} = 8.10 \) would be considered exceptionally bright, implying that its infrared emission is significant relative to typical \ac{AGN} hosts.

Regarding the colours:
\begin{itemize}
    
    \item The W1--W2 colour of \( -0.01 \pm 0.04 \) lies near zero, indicative of a source whose infrared SED is characteristic of a cool stellar population with minimal hot dust or \ac{AGN} torus contribution \citep{2010AJ....140.1868W}.
    
    \item The W2--W3 colour of \( 2.81 \pm 0.04 \) and W3--W4 colour of \( 2.01 \pm 0.05 \) show significant brightening at longer mid-IR wavelengths, consistent with warm dust emission from star-forming regions and/or lower-luminosity \ac{AGN} systems \citep{Cluver2014, Cluver_2017}.

\end{itemize}





\begin{figure}
     \centering
     \includegraphics[width=\linewidth]{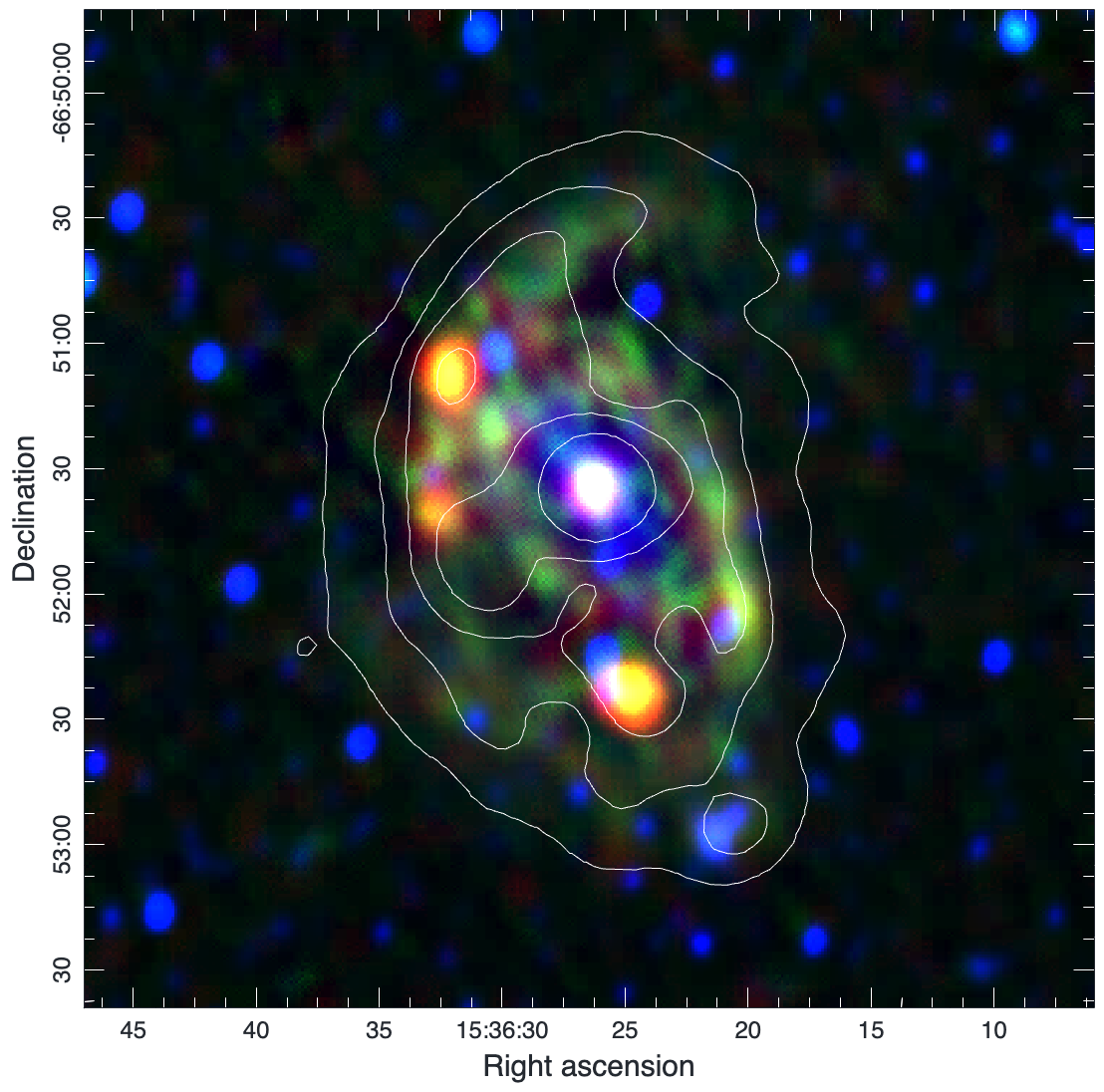}
     \caption{RGB \ac{WISE} Red-W4, Green-W3, and Blue-W1. W1 colour indicates excess hot dust emission from an \ac{AGN} while W3 and W4 luminosity is sensitive to star formation. Radio Contours 3, 9, 15, 25, and 50\,$\sigma$ applied from \ac{EMU} radio observations.}
     \label{fig_rgb4}
\end{figure}

The position of our target galaxy \ngc\ (black circle in Figure~\ref{fig_SF}-top panel) on the WISE colour–colour diagram indicates W1--W2 = $-0.01 \pm 0.04$ and W2--W3 = $2.81 \pm 0.04$. These values place the source below the AGN-dominated region but firmly within the dusty, star-forming locus, consistent with the regime of galaxies exhibiting significant warm dust emission \citep{Jarrett_2023}. The location also lies above the stellar-dominated sequence, suggesting that mid-infrared emission from heated dust makes a substantial contribution.
While its location on the \ac{SFMS} in Figure~\ref{fig_SF}-bottom panel confirms its classification as a star forming galaxy \citep{Brinchmann_2004}, where its star formation rate $\log \mathrm{SFR}\;[M_\odot\,\mathrm{yr}^{-1}] = 0.58 \pm 0.08$ is obtained from \citep{2024MNRAS.529.3210B} and the stellar mass $\log \rm M* = 11.24 \pm 0.10$ is taken from \citet{2018ApJ...864...40P}.
\begin{sloppypar}
The presence of cool dust around the centre of the \ngc, as indicated by W2--W3 colour, could be due to interactions between high-velocity particles in \ac{AGN} jets and the interstellar dust grains \citep{10.1093/mnras/stx403, Cluver2014}. However, it is important to note that in the W1-W3-W4 RGB image (Figure \ref{fig_rgb4}), we do not observe any increase in infrared emission, disturbance in dust distribution, or other signs of jet-\ac{ISM} interaction coincident with the radio jet's location. No other prominent infrared sources are detected within $6\arcsec$ of the radio jet. The native angular resolution of \ac{WISE} is $\sim 6\arcsec$ in W1--W3 and $\sim12\arcsec$ in W4 \citep{2010AJ....140.1868W}, which is sufficient to trace global dust emission but not to resolve small-scale interactions between the radio jet and the interstellar dust. Thus, the observed \ac{WISE} colours provide evidence for cool dust in the galaxy, although higher-resolution infrared data would be required to directly probe any jet–ISM interaction.
\end{sloppypar}

    \begin{table}[hbt!]
    \begin{threeparttable}
    \caption{Infrared Magnitude - \ac{WISE}}\label{tab_wise}
    \begin{tabular}{llll}
    \toprule
    WISE Bands & Magnitude (Vega) & Magnitude (AB) \\
    \midrule
    W1   &  8.10 $\pm$ 0.02 & 10.80 $\pm$ 0.02  \\
    W2   &  8.11 $\pm$ 0.04 & 11.45 $\pm$ 0.04  \\
    W3   &  5.30 $\pm$ 0.06 & 10.54 $\pm$ 0.06  \\
    W4   &  3.29 $\pm$ 0.08 & 9.89 $\pm$ 0.08  \\
    WI--W2    & $-0.01 \pm$ 0.04 & $-0.65 \pm$ 0.04\\
    W2--W3    &  2.81 $\pm$ 0.04 & 0.91 $\pm$ 0.04 \\
    W3--W4    &  2.01 $\pm$ 0.05 & 0.65 $\pm$ 0.05 \\
    \midrule
    \midrule
    Surface Brightness    & 17.437  \\
    $\log$ M*[M-solar]    & 11.24 $\pm$ 0.10 \\
    $\log$ SFR [M-solar/Year]    & 0.58 $\pm$ 0.08  \\
    \bottomrule
    \end{tabular}
    \end{threeparttable}
    \end{table}

 \begin{figure}
     \centering
     \includegraphics[width=0.9\columnwidth]{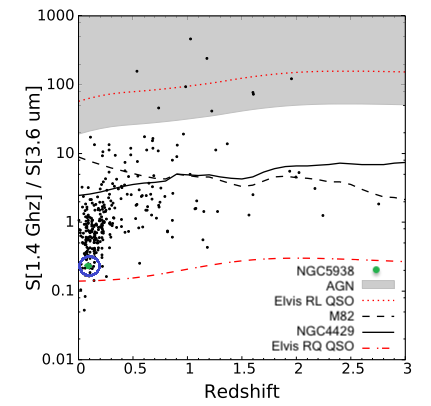}
     \caption{The ratio between the radio 1.4 GHz and extrapolated 3.6\,$\mu$m integrated flux density plotted as a function of redshift. Black points represent galaxies from \citet{2018MNRAS.473.4523W}. The green circled dot marks the flux density ratio of NGC 5938 (this work). The grey shaded area denotes the loci of radio-loud AGN as per \citet{2018MNRAS.473.4523W}. Template tracks are shown for the starburst galaxy M82 (black dashed line), and the spiral galaxy NGC 4429 (black solid line) from \citet{Seymour_2007}, radio-loud quasars (red dotted line), and radio-quiet quasars (red dash-dotted line) from \citet{1994ApJS...95....1E}.
     }
     \label{fig_ratio}
 \end{figure}

\subsubsection{Infrared and Radio Emission Analysis}\label{subsec4}

To further investigate \ac{AGN} activity in \ngc, we calculated the radio-to-infrared flux density ratio, a valuable diagnostic tool commonly used in identifying \ac{AGN} jets \citep{2018MNRAS.473.4523W}. This ratio was computed using the integrated radio flux density at 1.4\,GHz (Table \ref{tab_Rflux}) and infrared flux density at 3.6\,$\mu$m to generate the ratio. We used the integrated radio flux density at 944\,MHz, $S_{944} = 90.6 \pm 0.49$~mJy (Table~\ref{tab_Rflux}), and the measured spectral index $\alpha = -1.5 \pm 0.2$ to extrapolate flux density at 1.4\,GHz using the power-law relation (Equation \ref{Eq1}), yielding $S_{1.4} = 50.16 \pm 4.27$~mJy.

\begin{equation}\label{Eq1}
    S_2 = S_1 \left( \frac{\nu_2}{\nu_1} \right)^{\alpha}
\end{equation}

For infrared measurements, the \ac{WISE} W1 Vega magnitude of $M_{W1} = 8.10 \pm 0.02$ was converted to a flux density \citep[Equation \ref{Eq2};][]{2010AJ....140.1868W}, where $S_{\nu_0} = 306.7$\,Jy is the W1 zero-magnitude flux density \citep{2010AJ....140.1868W}. This yields $S_{3.4} = 176.48 \pm 4.0$\,mJy. Extrapolating to $3.6\,\mu$m using the measured infrared spectral index ($\alpha_{\rm IR} \approx -1.0$) gives $S_{3.6} \approx 209.5 \pm 4.0$\,mJy.

\begin{equation}\label{Eq2}
     S_{W1} = S_{\nu_0} \times 10^{-0.4 \times M_{W1}}
\end{equation}  

\begin{equation} \label{Eq3}
R_{\rm 1.4/3.6} = \frac{S_{1.4,{\rm GHz}}}{S_{3.6,\mu{\rm m}}}
\end{equation}

The radio-to-infrared ratio (Equation~\ref{Eq3}) is $0.24 \pm 0.02$, which lies well below the canonical radio-loud quasar threshold ($R > 10$; \citealt{1994ApJS...95....1E}), indicating that radio emission is not the dominant power source. In Figure~\ref{fig_ratio}, \ngc\ (green dot) lies marginally above the median star-forming galaxy locus but well below the radio-loud \ac{AGN} regime.

To place this ratio in a broader context, we compared \ngc’s stellar mass ($\log M_* / M_\odot = 11.02 \pm 0.05$) and radio luminosity at 888 MHz ($\log L_{888{\rm MHz}} = 21.50 \pm 0.03$\,W\,Hz$^{-1}$) to the distributions of \ac{AGN} hosts identified using optical \ac{BPT} diagnostics, \ac{WISE} colours, the Mass–Excitation diagram, and ProSpect(0010) modelling from \citet{2024PASA...41...16P}. While the radio-to-infrared ratio alone would not classify \ngc\ as strongly radio-loud, its radio luminosity is consistent with the median values for \ac{AGN} selected by ProSpect(0010), a method designed to detect lower-luminosity \ac{AGN} in massive hosts. 

This quantitative cross-comparison indicates that \ngc\ likely harbours a low luminosity \ac{AGN} where radio output is modest compared to its infrared emission, though consistent with multi-wavelength \ac{AGN} diagnostics. The results emphasise that in composite systems, where star formation and \ac{AGN} activity both contribute to the observed flux, a single diagnostic can under-represent the \ac{AGN} component, and multi-method analyses are required for reliable classification.

\subsection{X-ray Observations}
 \label{subsec5}

\subsubsection{X-ray emission from the radio jet}

\begin{figure}
     \centering
     \includegraphics[width=\linewidth]{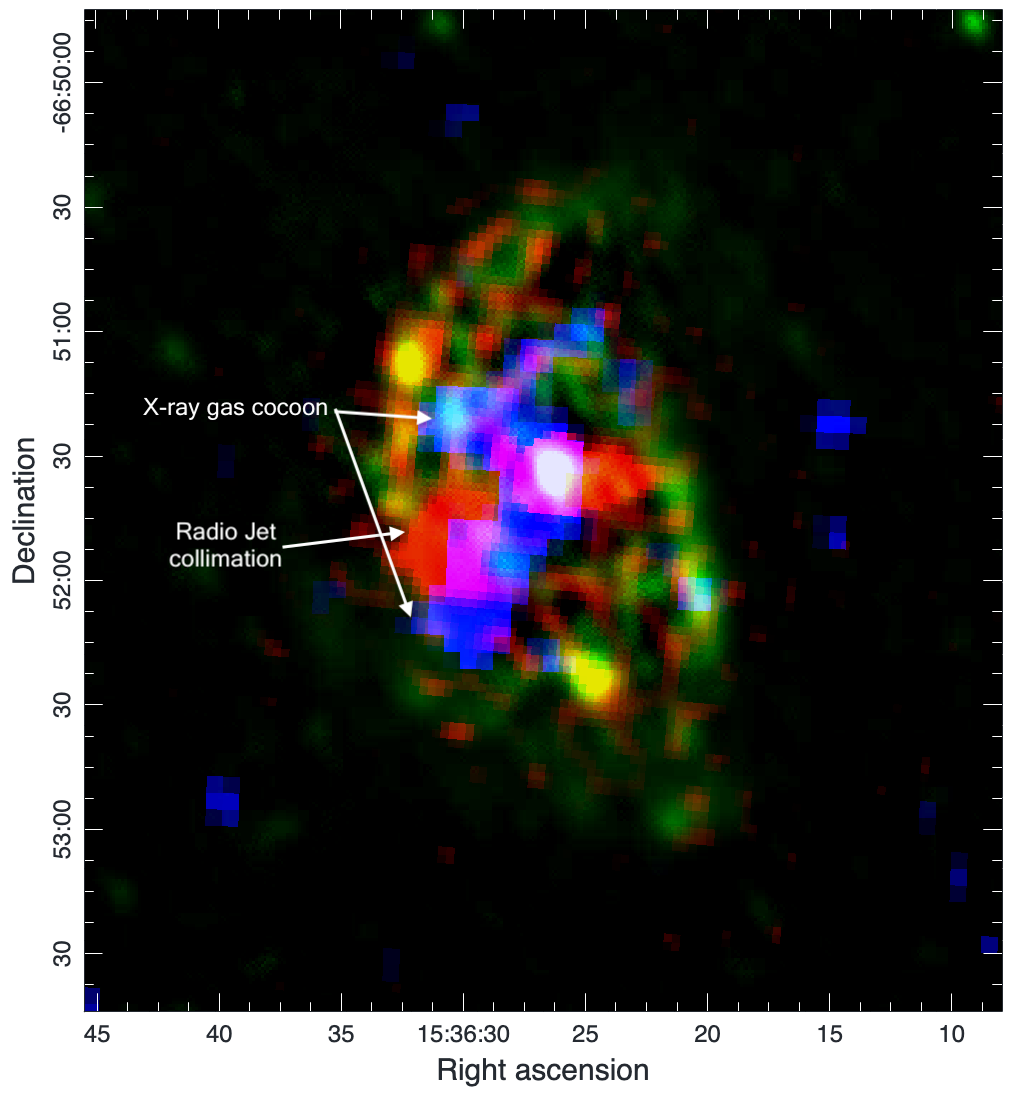}
     \caption{RGB composite image of \ngc\ illustrating the distribution of radio emission (Red)-\ac{EMU} Hi-Res, infrared W3 emission (Green)-\ac{WISE} and X-ray emission (Blue) - \ac{eROSITA}. The radio jet seems to be collimated by the surrounding X-ray gas.}
     \label{fig_rgb3}
 \end{figure}

In the X-ray regime, we investigated the galaxy \ngc\ utilising the \ac{eROSITA}-eRASS:4 X-ray 0.2--5.0\,keV processed image (Figure~\ref{fig_mul}d). Using the \texttt{ermldet} task, no strong point-like X-ray source was detected at the radio jet’s position.  Instead, the observed X-ray morphology reveals extended, diffuse emission broadly following the radio contours (Figure~\ref{fig_mul}d, Figure~\ref{fig_rgb3}), consistent with a cocoon of shocked gas surrounding the jet. To quantify this, we extracted photons within the $3\sigma$ radio contour, detecting 29 counts (4 background) in the 0.2–2.3 keV band, corresponding to a significance $> 3\sigma$.

To estimate the X-ray upper limit of the jet region,  we extracted X-ray spectra using \texttt{srctool} with a $10 \arcsec$ circular region centred at the position of the east radio jet (EEF $=0.4$; \citealt{tubin-arenas_etal2024}) and an annulus ($70{\arcsec}$--$130{\arcsec}$) for background subtraction. Following \citet{kraft_etal1991}, assuming a power-law spectrum with galactic absorption ($N_{\mathrm{H}} = 1.4 \times 10^{21}\ \mathrm{cm}^{-2}$) and photon index $\Gamma = 1.8$, we obtained a $3\sigma$ flux upper limit of $S_{0.2 - 2.0\,\mathrm{keV}} \approx 0.7 \times 10^{-13}\ \mathrm{erg\ cm^{-2}\ s^{-1}}$, resulting in a soft X-ray luminosity of $L_{\mathrm{X}} \approx 6.0 \times 10^{39}\,\mathrm{erg\,s^{-1}}$ ($\log L_{\mathrm{X}} \approx 39.8$).

To assess the possibility of a central \ac{AGN}, we extracted the X-ray events from a $10\arcsec$ circular region centred at the core (Figure~\ref{fig_rgb3}). This yielded 7 photons (0.7 background), significant at $> 3\sigma$. However, all photons fall below ($2$ keV), with no hard X-rays ($\geq 2\,\mathrm{keV}$) detected. As a result, the inferred luminosity pertains only to the soft X-ray band. Due to the low photon counts, we estimated the soft X-ray flux assuming an absorbed power-law model ($N_{\mathrm{H}} = 1.4 \times 10^{21}\ \mathrm{cm}^{-2}$ and $\Gamma = 1.8$). The resulting soft X-ray luminosity ($3.4\pm3 \times 10^{39}\,\mathrm{erg\,s^{-1}}$) is consistent with LINER galaxies ($\log L_{\mathrm{X}} \approx 39.9$) and higher than H\textsc{ii} galaxies ($\log L_{\mathrm{X}} \approx 39.3$), while remaining well below the typical Seyfert regime \citep[$\log L_{\mathrm{X}} \approx 41.0$,][]{Georgantopoulos1999}. Notably, this luminosity is also broadly consistent with empirical relations for X-ray binary (XRB) populations and shock-heated plasma in a galaxy with an SFR $\approx 0.58$ $M_{\odot}$/yr \citep{10.1111/j.1365-2966.2011.19862.x, Ranalli2003}. This suggests that the observed soft emission may be dominated by stellar processes and shock-heated gas associated with the radio-jet interaction rather than the \ac{AGN} itself. 

 Although no hard-band emission is detected,  we evaluated whether \ngc\ could host a heavily obscured near-Compton-thick \ac{AGN} \citep{2015ApJ...815L..13R}. We derived $3\sigma$ upper limits from the \ac{eROSITA} data, assuming a heavily obscured power-law model with $N_{\rm H} \gtrsim  1.0 \times 10^{24}\,\mathrm{cm^{-2}}$ and $\Gamma = 1.8$. This observed hard-band upper limit corresponds to an intrinsic luminosity limit of $L_{2-10\,\mathrm{keV}} \lesssim 6.8 \times 10^{42}\,\mathrm{erg\,s^{-1}}$, with a corresponding soft-band intrinsic limit of $L_{0.2-2\,\mathrm{keV}} \lesssim 6.6 \times 10^{42}\,\mathrm{erg\,s^{-1}}$. These limits, along with the estimated soft X-ray luminosity under low absorption scenario, imply that NGC~5938 does not host a luminous Seyfert-like \ac{AGN}, and that any nuclear activity is either intrinsically weak or heavily absorbed \citep{Gilli2007, Brightman2012, 2015ApJ...815L..13R}. Comparison with the BASS survey \citep{Tokayer2025} further shows that \ngc\ is significantly less luminous and lacks the detectable hard X-ray emissions ($\log L_{14-195\,\mathrm{keV}} \geq 43$), characteristics of the powerful \ac{AGN} probed in the study \citet{Tokayer2025}. 

Although more complex absorption scenarios, such as highly ionised absorbers, could in principle suppress the hard band while allowing some soft emission \citep{Done_2007}; however the limited signal-to-noise ratio of the available data prevents any meaningful constraints on these models. These results are consistent with a scenario in which the \ac{AGN} in \ngc\ may be undergoing a low-accretion-rate, \emph{pre-blowout} phase, where the nuclear feedback is currently insufficient to clear the surrounding obscuring material. We therefore conclude that the current data cannot rule out a deeply buried, low-luminosity nucleus, although the observed emission is more naturally explained by non-AGN processes. Deeper hard X-ray observations (e.g. with \emph{NuSTAR}; \citealt{Harrison2013}) would be required to detect or place meaningful constraint on any obscured nuclear emission.


\subsubsection{X-Ray and Radio Emission Interplay} 
Focusing on the centre of the galaxy's nucleus, the X-ray continuum image (Figure~\ref{fig_mul}d) 
reveals emission concentrated towards the galaxy's nucleus, as further delineated by the overlaid radio contours. The observed X-ray jet-like emission could originate from the superheated gas spiralling towards a supermassive black hole or outflows often associated with \ac{AGN} activity \citep{book2}. However, the relationship between the extended X-ray gas and radio halo is complex. While at a glance, the multi-frequency composite (Figure~\ref{fig_rgb3} radio emission is shown in red, X-ray emission in blue) image might visually suggest a less direct spatial overlap, a closer examination (Figure~\ref{fig_mul}d inset) reveals that the brighter X-ray emission is primarily adjacent to and often outlines the radio contours, especially towards the south-east. This diffuse and enveloping morphology indicates that the radio jet is likely interacting with, and potentially confined by, the more extended X-ray gas/halo.

Similar to the systems like ESO~295--IG022 \citep{2001A&A...369..467R}, Centaurus\,A \citep{1982IAUS...97..107F}, UGCA\,127 \citep[Phaedra,][]{2025ApJ...984L..52F} and NGC\,1068 \citep{Young_2001}, where jet-\ac{ISM} interactions are prominent, the X-ray emission in \ngc\ might be indicative of diffuse, shocked gas at the jet's edge and along its boundaries.
While Centaurus\,A exhibits X-ray cavities, it also shows significant X-ray emissions from shocks at the edge of its lobes.  Similarly NGC\,1068 displays X-ray features linked to direct jet-cloud impacts and shocks. Therefore, in the case of \ngc, in addition to potentially larger-scale confinement or channelling, the observed X-ray morphology suggests the possibility of shocked emission due to the jet's interaction with the surrounding medium at the boundary. This interaction could be similar to the ``cocoon`` effect seen in ESO~295--IG022, where the jets may be collimated by the surrounding X-ray gas, maintaining pressure balance between the central \ac{ISM} and the gas within the jet. This would also be very suggestive of channelling effects taking place, whereby the violent radio continuum jets can punch holes and displace the X-ray emitting cluster gas, as is also seen in the NGC\,1275 jet at the centre of the Perseus cluster \citep{2021ApJ...911...56G}.

\section{Conclusion} 
\label{S:Conclusion}

Here we presented a multi-frequency study of the barred spiral galaxy \ngc\ (Araish), leveraging new \ac{EMU} and \ac{RACS} surveys data. Our analysis reveals the presence of extended, steep spectrum radio emission with a spectral index of $\alpha = -1.2 \pm 0.2$ suggesting synchrotron emission indicative of an \ac{AGN} jet. The radio jet extends perpendicular to the galaxy's major axis, reaching a physical size of $\approx 8.2$\,kpc. The galaxy's nucleus has a spectral index ($\alpha = -0.7 \pm 0.4$), consistent with the signature of core \ac{AGN} activity \citep{Stren2012}. 

The host galaxy is found to lie on the star-forming main sequence, and \ac{WISE} colour diagnostics place it firmly within the star-forming regime, with no indication of a classical quasar-like \ac{AGN}. However, the radio-to-infrared flux ratio places \ngc\ in a region occupied by radio-quiet quasars, this finding is critically consistent with the recognised continuum of radio activity, in which many systems exhibit composite radio output from low luminosity jets and/or compact cores together with star-formation-related emission i.e. “radio-intermediate” or low-excitation/low-luminosity \ac{AGN} \citep{2012MNRAS.421.1569B}. 

The X-ray properties of NGC 5938 provide further insight into this composite nature. X-ray emission from the galaxy's nucleus aligns with the radio contours, suggesting a concentrated source. The soft X-ray luminosity ($L_X \approx 3.4 \times 10^{39} erg s^{-1}$) is consistent with that expected from X-ray binary populations and shock-heated plasma for the galaxy's star-formation rate (SFR $\approx 0.58$ $M_{\odot}$/yr). While the absence of hard X-ray emission ($\geq 2$ keV) disfavours a luminous, unobscured AGN, our analysis of the 2–10 keV upper limits suggests that a deeply buried or low-accretion-rate nucleus cannot be ruled out. Such a system may be in a `pre-blowout' phase, where nuclear feedback is insufficient to clear the surrounding obscuring material. Crucially, the observed spatial relationship, where the X-ray outlining follows the extended radio jet (similar to systems like ESO~295-IG022, Centaurus A, and NGC~1068), is highly suggestive of jet confinement and channelling effects caused by the surrounding X-ray gas or halo. This displacement of the X-ray emitting gas by the radio jets, a phenomenon seen in systems such as NGC~1275, supports a scenario where the jet is being collimated. 

The barred spiral morphology of \ngc\ likely contributes to the funnelling of gas towards the central black hole \citep{10.1093/mnras/stv235}, enhancing accretion and powering the observed \ac{AGN} activity. While the optical morphology suggests possible interaction of the jet with the host galaxy’s interstellar medium, current infrared data provide no direct evidence of strong jet–ISM coupling. 

It is critical to acknowledge that identifying \ac{AGN} and disentangling their contributions from star formation remains a significant challenge. Different selection methods can reveal \ac{AGN} populations with varying characteristics, introducing potential selection biases. This interplay remains tentative and will require multi-band follow-up to confirm. To fully constrain the physical parameters driving jet confinement, future work should prioritize higher-resolution radio observations and deeper hard X-ray spectral modelling (e.g., with NuSTAR) to accurately map the thermal and pressure structure of the surrounding medium.

The combination of an active central engine producing a synchrotron jet and a disk galaxy with ongoing star formation firmly establishes \ngc\ as a compelling new example of a spiral \ac{DRAGN}. It represents a crucial laboratory for studying the effects of \ac{AGN}-driven jets in non-elliptical hosts, offering unique insights into feedback mechanisms in disk-dominated galaxies. This case study highlights the power of the \ac{EMU} survey in uncovering rare, unusual galaxies when combined with multi-wavelength ancillary data, and underscores the potential for future systematic searches to identify further spiral \ac{DRAGN}s and constrain their role in galaxy evolution.


\begin{acknowledgement}
This scientific work uses data obtained from Inyarrimanha Ilgari Bundara/the Murchison Radio-astronomy Observatory. We acknowledge the Wajarri Yamaji People as the Traditional Owners and native title holders of the Observatory site. CSIRO’s ASKAP radio telescope is part of the Australia Telescope National Facility (https://ror.org/05qajvd42). Operation of ASKAP is funded by the Australian Government with support from the National Collaborative Research Infrastructure Strategy. ASKAP uses the resources of the Pawsey Supercomputing Research Centre. Establishment of ASKAP, Inyarrimanha Ilgari Bundara, the CSIRO Murchison Radio-astronomy Observatory and the Pawsey Supercomputing Research Centre are initiatives of the Australian Government, with support from the Government of Western Australia and the Science and Industry Endowment Fund. 

This work is also based on data from eROSITA, the soft X-ray instrument aboard SRG, a joint Russian-German science mission supported by the Russian Space Agency (Roskosmos), in the interests of the Russian Academy of Sciences represented by its Space Research Institute (IKI), and the Deutsches Zentrum für Luft- und Raumfahrt (DLR). The SRG spacecraft was built by Lavochkin Association (NPOL) and its subcontractors, and is operated by NPOL with support from the Max-Planck Institute for Extraterrestrial Physics (MPE). The development and construction of the eROSITA X-ray instrument was led by MPE, with contributions from the Dr. Karl Remeis Observatory Bamberg, the University of Hamburg Observatory, the Leibniz Institute for Astrophysics Potsdam (AIP), and the Institute for Astronomy and Astrophysics of the University of Tübingen, with the support of DLR and the Max Planck Society. The Argelander Institute for Astronomy of the University of Bonn and the Ludwig Maximilians Universität Munich also participated in the science preparation for eROSITA. The eROSITA data shown here were processed using the eSASS/NRTA software system developed by the German eROSITA consortium.

M.D.F. and S.L. acknowledge Australian Research Council (ARC) funding through grant DP200100784. 

T.A. acknowledges the support from the National SKA Program of China (2022SKA0120102),  FAST special funding (NSFC 12041301), and the Xinjiang Tianchi Talent Program.

\end{acknowledgement}








\bibliography{references}


\end{document}